
\input phyzzx

\def\bpsi {\bar\psi}
\def\der {\partial}
\def\bder {\bar\partial}
\def\lch #1#2{L_{#1}^{(#2)}}
\def\blch #1#2{\bar L_{#1}^{(#2)}}
\def\fr #1#2{{#1\over #2}}
\def\bz {\bar z}
\def\lo {\bz\bder-z\der}
\def\lkts {L_{-k}^{(2s)}}
\def\blkts {\bar L_{-k}^{(2s)}}
\def\loi {\bz_i\bder_i-z_i\der_i}
\def\cL {{\cal L}}

\Pubnum={IFT - P.001/93}
\titlepage
\title{Off - critical $W_\infty$ and Virasoro Algebras As Dynamical
Symmetries Of The Integrable Models\foot{
Talk given at the NATO Workshop ``Integrable Quantum Field Theories'',
Como, Italy, September 1992.}}
\author{Galen Sotkov\foot{
On leave of absence from the Institute of Nuclear Research
and Nuclear Energy, Sofia, Bulgaria.}}
\address{Instituto de F\' isica Te\' orica, Universidade Estadual Paulista,
Rua Pamplona 145, 01405-900 - S\~ ao Paulo, S.P., Brazil}
\author{Marian Stanishkov\footattach}
\address{Instituto de F\' isica, Universidade de S\~ ao Paulo,
Caixa Postal 20516, 01498 - S\~ ao Paulo, S.P., Brazil}
\abstract{
We find an infinite set of new noncommuting conserved charges in a
specific class of perturbed CFT's and  present
a criterion for their existence.
They appear to be higher momenta of the already known commuting conserved
currents.  The
algebra they close
consists of two noncommuting $W_\infty$ algebras. We find various Virasoro
subalgebras of the full symmetry algebra. It is shown on the examples of
the perturbed Ising and Potts models
that one of them
plays an essencial role in the computation of the correlation
functions of the fields of the theory.}
\endpage

\null
\voffset-1cm
\chapter {Noncommuting conserved charges of IM's}

 The explicit construction of all the integrals of motion (conserved
charges) for a given dynamical problem allows to solve it exactly. An
important tool in the realization of such a program is the algebra of the
conserved charges and its representations. The power of this {\it
symmetries} strategy to the problem of the exact solution of $2-D$
integrable models ( IM's ) was demonstrated by the recent development of
the $2-D$ conformaly invariant theories
\REF\bpz{A.A. Belavin, A.M. Polyakov, A.B. Zamolodchikov\journal
Nucl.Phyz.&B241(84)333.}
[\bpz]. The key point in the
construction of the correlation functions of the fields of this class of
IM's is the appearence of the Virasoro algebra:
$$
[\cL_n,\cL_m]=(n-m)\cL_{n+m}+{c\over 12}n(n^2-1)\delta_{n+m}
\eqn\vir
$$
and its generalizations - $SU(N)$ - Kac-Moody, super - Virasoro, $W_N$,
$W_\infty,\ldots$, as symmetries of the model. In all these cases the
generators of the corresponding algebras are realized as higher momenta of
the conserved currents. For example $\cL_n$'s are all the momenta of the
stress - tensor $T_{\mu\nu}=(T,\bar T)$:
$$
\eqalign{
    \cL_n &=\oint dz z^{n+1}T(z), \qquad\qquad
\bar \cL_n=\oint d\bar z\bar z^{n+1}\bar T(\bar z)\cr }.
$$

Since the conformal models belong to the big family of the relativistic
integrable models one could wonder whether analogous {\it infinite
symmetries algebra} approach works in the case of the {\it nonconformal
IM's}, say - sin-Gordon, massive fermions, affine Toda models etc. As it is
known, the integrability of all these models is based on the existence of
an infinite set of conserved charges ( CC ):
$$
\eqalign{
P_s &=\oint T_{2s}dz-\oint \Theta_{2s-2}d\bar z,\qquad\qquad
\bar P_s=\oint\bar T_{2s}d\bar z-\oint\Theta_{2s-2}dz\cr
T^{\mu_1\ldots\mu_{2s}} &=(T_{2s}, \bar T_{2s}, \Theta_{2s-2}),\qquad
\bar\partial T_{2s}=\partial \Theta_{2s-2}\qquad
\partial\bar T_{2s}=\bar\partial\Theta_{2s-2}\cr}
\eqn\pss
$$
they have. However, the algebra of the $P_s$ ($\bar P_s$) is {\it abelian}:

\noindent
$
[P_s,P_{s'}]=0=[P_s,\bar P_{s'}]
$
and this is considered to be an obstacle in using these symmetries for the
calculation of the exact Green functions of the model. Therefore, the
question one has to answer first is {\it whether} $P_s$ exhaust all the
conserved charges of these models,{\it i.e.}

1) are there more (nontrivial) conservation laws?

2) if so, is it the algebra of the new conserved charges {\it nonabelian} ?

Addressing such a question we already have a hint that the answer should be
positive ( at least for a certain class of IM's ). It is the {\it important
observation} of Thacker and Itoyama
\REF\ti{H. Itoyama, H.B. Thacker\journal Nucl.Phys.&B320(89)541}
[\ti] that the $XY$ - model in high
temperature phase ( {\it i.e.} {\it massive} Dirac fermion ) possesses on
top of the abelian charges \pss\ a large set of noncommuting CC's.
They close two different off - critical Virasoro algebras. The
best way of generalizing this result is to find the geometrical origin of
these off- critical symmetries and to describe the corresponding class of
models having this symmetry. Being far from the full understanding of
these infinite
symmetries of the IM's we choose to follow the more practical way of
explicit construction of the noncommuting CC's. Our starting
point is the fact that allmost all relativistic (~nonconformal ) IM's can
be represented as an appropriate perturbation of certain conformal models
\REF\zam{A.B. Zamolodchikov\journal Int.Journ. of Mod.Phys.&A3(88)743;
in Adv.Studies of Pure Math., vol.19 (1989)641.}
[\zam], {\it i.e.}:
$$
S_{IM}=S_{conf}+g\int\Phi_\Delta(z)\bar\Phi_\Delta(\bar z)d^2z .
\eqn\pert
$$
This suggests that the desired new charges (if they exist) should be
realized as specific combinations of the higher momenta of the conserved
tensors ( $T_{2s}$, $\bar T_{2s}$, $\Theta_{2s-2}$ ):
$$
\eqalign{
{\cal F}^{(n)}_{2s} &=\sum_{k=1}^s\bigg\{ \alpha _k(g)z^{2k-1+n}
\bar z^{\gamma (k,n)}T_{2k}(z,\bar z) + \bar\alpha _k(g)\bar z^{2k-1+n}
z^{\bar\gamma (k,n)}\bar T_{2k}(z,\bar z)\cr
+ &\beta _k(g)z^{\delta (k,n)}\bar z^{\bar\delta (k,n)}
\Theta _{2k-2}(z,\bar z)\bigg\} \cr } ,
\eqn\mom
$$
such that
$$
\bar\partial{\cal F}^{(n)}_{2s}=\partial{\cal G}^{(n)}_{2s-2} ,
$$
where
$
\alpha_k(g)=\alpha_k g^{{s-k\over 1-\Delta}},\qquad
\beta_k(g)=\beta_k g^{{s-k\over 1-\Delta}} .
$

The {\it crucial observation} that simplifies the construction of the new
conservation laws ${\cal F}_{2s}^{(n)}$ is the following {\it criterion
of existing of such quantities}: {\it If the conservation laws of the spin
- 2s} ( $s>1$ ) {\it tensors} $T_{2s}$ {\it are in the form}:
$$
\eqalign{
\bar\partial T_{2s} &=\partial^{2s-1}\Theta+g^p\sum_{l=1}^{s-1}
A_l\partial^{2(s-l)-1}T_{2l}\cr
\partial\bar T_{2s} &=\bar\partial^{2s-1}\Theta+g^p\sum_{l=1}^{s-1}
A_l\bar\partial^{2(s-l)-1}\bar T_{2l}\cr
\bar\partial T &=\partial\Theta,\qquad\qquad\partial T=\bar
\partial \Theta,\qquad\qquad p={1\over 1-\Delta},\cr}
\eqn\claws
$$
{\it then there exist} $4s-3$ {\it new conserved currents} ${\cal F}_{2s}^{
(n)}\qquad (n=1,2,\ldots,4s-3)$ {\it for each fixed} $s=1,2,\ldots$ . In
words the existence of new conserved charges:
$$
\lch {-n}{2s}=\int{\cal F}_{2s}^{(n)}dz-
\int{\cal G}_{2s-2}^{(n)}d\bar z\ \qquad
\bar L_{-n}^{(2s)} =\int\bar{\cal F}_{2s}^{(n)}-
\int\bar{\cal G}_{2s-2}^{(n)}dz
\eqn\lns
$$
{\it is hidden in the specific form of the traces} $\Theta_{2s-2}$ {\it of
the traditional conserved currents} $T_{2s}$:
$$
\Theta_{2s-2}=\partial^{2s-2}\Theta+g^p\sum_{l=1}^{s-1}A_l
\partial^{2(s-l)-2}T_{2l} .
$$

Instead of general proof we shall show here how our criterion is working on
the simplest example.  Suppose we have a model such that:
$$
\bar\partial T=\partial\Theta ,\qquad\qquad \partial\bar T=
\bar\partial\Theta
$$
and
$$
\bder T_4 =\der^3\Theta+\beta g^2\der T ,\qquad
\der \bar T =\bder^3\Theta+\beta g^2\bder \bar T .
\eqn\tfour
$$

We are going to demonstrate that on top of the usual commuting charges:
$$
\eqalign{
L_{-1} &=\int Tdz-\int \Theta d\bar z,\qquad \bar L_{-1}=
\int \bar Td\bar z-\int\Theta dz\cr
L_{-3}^{(4)} &=\int T_4dz-\int\Theta_2d\bar z,\qquad
\bar L_{-3}^{(4)}=\int\bar T_4d\bar z-\int \Theta_2dz\cr}
$$
and the Lorentz rotation generator
$$
L_0=\int(zT+\bar z\Theta)dz-\int(\bar z\bar T+z\Theta)d\bar z
$$
five new conserved charges appear. Let us consider the ``nonconservation
laws'' of the following three quantities: ${\bar z}^2\bar T$, $z\bar z T$
and $z^2T_4$:
$$
\eqalign{
\der(\bar z^2\bar T) &=\bder(\bar z^2\Theta)-2\bar z\Theta\cr
\bder(z\bar z T) &=zT +\der(z\bar z\Theta)-\bar z\Theta\cr
\bder(z^2T_4) &=\der^3(z^2\Theta)-6\der(z\der\Theta)+\beta
g^2\der(z^2T)-2\beta g^2zT\cr}.
$$

A simple algebra leads to the following new conservation law:
$$
\bder{\cal F}_4^{(1)}=\der\tilde\Theta_2^{(1)},
\eqn\newcl
$$
where
$$
\eqalign{
{\cal F}_4^{(1)} &=z^2T_4+2\beta g^2z\bar zT+\beta g^2\bar z^2\Theta\cr
\tilde\Theta_2^{(1)} &=\beta g^2(z^2T+\bar z^2\bar T)+
2\beta g^2z\bar z\Theta -6z\der\Theta+\der^2(z^2\Theta)\cr}
\eqn\ftheta
$$
In the same way one can get the ``first momenta'' of $T_4$ conservation
law:
$$
L_{-2}^{(4)}=\int(zT_4+\beta g^2\bar zT)dz-
\int(\beta g^2zT+g^2\bar z\Theta)d\bar z .
\eqn\fmomtfour
$$
The remaining three conserved charges are the ``complex conjugated'' of
$L_{-1}^{(4)}$ and $L_{-2}^{(4)}$ and the ``third momenta'' of $T_4$
conservation law $L_0^{(4)}$ including $z^3T_4$, ${\bar z}^3T_4$,
$z^2\bar z T$ etc.


In principle one can repeat this procedure for all the $T_{2s}$'s and to
construct an infinite set of new conserved charges: $L_{-n}^{(2s)}$,
${\bar L}_{-n}^{(2s)}$, $n=0,1,\ldots ,2s-1$.

\chapter {Symmetries of the off - critical Ising model}

Turning back to our problem of {\it constructing noncommuting conserved
charges} for the IM's given by \pert\ we have to check whether exist models
which satisfy our criterion, {\it i.e.} their standard $T_{2s}$ -
conservation laws to be in the form \claws\ .

The simplest case is the set of models obtained by $\Phi_{\Delta_{1,3}}
\bar\Phi_{\Delta_{1,3}}$ perturbations of the conformal minimal models
( $c_p=1-{6\over (p+1)(p+2)}$, $\Delta_{1,3}(p)={p\over p+2}$ )
(see [\zam]).
They have all the $T_{2s}$, $\bar T_{2s}$, $s=1,2,\ldots$ conserved. The
first model ($p=2$) of this set is the thermal perturbation of the Ising
model which in the continuum limit coincides with the theory of free massive
Majorana fermion ( $\psi$, $\bar\psi$ ):
$$
\bder \psi=m\bar\psi\qquad\qquad \der\bpsi=-m\psi
\eqn\eqm
$$
$$
T={1\over 2}\psi\der\psi ,\qquad \bar T={1\over 2}\bar\psi\bder\bar\psi ,
\qquad \Theta=m\bar\psi\psi .
\eqn\tpsi
$$

To find the explicit form of $\Theta_{2s-2}$ in this case is better to use
the equation of motion \eqm\ instead of the conformal perturbative
technics. The corresponding conservation tensors of spin
$2s$ can be taken in the form
$
T_{2s}=\psi\partial^{2s-1}\psi ,\quad s=2,3,\ldots .
$

Simple computations based on the eq. \eqm\ leads to the following desired
form of the $T_{2s}$ - conservation laws:
$$
\eqalign{
\bder T_4 &=\der^3\Theta +2m^2\der T\cr
\bder T_6 &=\der^5\Theta+m^2 (\der T_4+4\der^3 T)\cr
\bder T_8 &=2\der^7\Theta+m^2(\der T_6+3\der^3T_4+2\der^5T)\cr}
\eqn\isclaw
$$
etc.
The conclusion is that this model satisfies our criterion
and therefore it
has $4s-3$ new conservation laws for each $s=1,2,\ldots$.
The corresponding conserved charges $L_{-n}^{(2s)}$, $\bar L_{-n}^{(2s)}$,
$0\le n\le 2s-1$, can be derived by the method we have demonstrated above
on the example of $L_{-1}^{(4)}$ (see eqs. \newcl\ ,\ftheta\ ).
For $s=2$ we have
together with the ``commuting charges'':
$$
\lch {-3}{4} =\int T_4dz-\int(\der^2\Theta+2m^2T)d\bar z ,\qquad
\blch {-3}{4} =\int\bar T_4d\bar z-\int(\bder^2\Theta+2m^2\bar T)dz
$$
the new charges $\lch {-2}{4}$,
$\lch {-1}{4}$ given by eqs. \ftheta\ and \fmomtfour\ with
$\beta =2$ and $g=m$, their ``conjugated''  $\blch {-2}{4}$,
$\blch {-1}{4}$ ( obtained from \ftheta\ and
\fmomtfour\ by the interchange
$z\leftrightarrow \bar z$, $T\leftrightarrow \bar T$,
$T_4\leftrightarrow\bar T_4$ ) and the unique $\lch {0}{4}$:
$$
\eqalign{
\lch {0}{4} &=\int{\cal F}^{(0)}_4dz-\int\Theta^{(0)}_4d\bar z\cr
{\cal F}^{(0)}_4 &=z^3T_4+6m^2z^2\bz T+2m^2\bz ^3\bar T+3m^2z\bz ^2
\Theta +\bder^2\bz ^3\Theta-9\bz ^2\bder\Theta\cr
\Theta^{(0)}_4 &=\bz^3\bar T_4+\ldots\equiv\bar {\cal F}^{(0)}_4 .\cr}
\eqn\newch
$$
The CC's arising from the $T_6$, $\bar T_6$ - conservation
laws \isclaw\ are given by:
$$
\eqalign{
\lch {-5}6 &=\int T_6dz-
\int\left[2\der^6\Theta+m^2(T_6+3\der^2T_4+2\der^4
T)\right]d\bar z\cr
\lch {-4}6 &=\int (zT_6+m^2\bz T_4)dz-\int (m^2zT_4+2m^4\bz T-4m^2\der T +
m^2\bz\Theta-\der^3\Theta+z\der^4\Theta )d\bz \cr }
$$
etc.
Following this method one could keep constructing $\lch {-n}{2s}$ for
higher spins. However, such a form of the conserved charges is inconvenient
both for deriving ( or guessing) the general form of $\lch {-n}{2s}$ and
for computing their algebra as well. For these purposes is better to have
$\lch {-n}{2s}$'s as differential operators acting on $\psi$ and
$\bar\psi$. One can do this in few steps. Starting from eqs.
\ftheta\ ,\fmomtfour\ ,\newch\
etc., we first exclude the time derivatives $\der_t\psi$ and then take
$t=0$ ( $z=t+x$, $\bar z=t-x$, $\der=\der_x$ ):
$$
\eqalign{
L_0 &=\fr 12\int x\left[ \psi\der\psi-\bpsi\der\bpsi+2m\bpsi\psi\right]
dx ,\qquad
L_{-1} =-\fr 12\int\left( \psi\der\psi +m\bpsi\psi\right) dx\cr
\lch {-1}4 &=\int\left\{ x^2\left[ \psi\der^3\psi+m\bpsi\der^2\psi-m^2\psi
\der\psi-m^3\bpsi\psi\right] +mx\bpsi\der\psi+2m\bpsi\psi\right\} dx\cr },
\eqn\xrep
$$
etc. The next step is to derive the momentum space form of $\lch {-n}{2s}$
by substituting the standard
creation and annihilation operators $a^{\pm}(p)$
decomposition of $\psi$ and $\bar\psi$:
$$
\eqalign{
L_0 &=\fr 12\int\fr {dp}{2\pi}p_0\left[(\der a^+)a^-
+(\der a^-)a^+\right] ,\qquad
\lch {-3}4 =\fr 14\int\fr {dp}{2\pi}\left( p_0-p\right)^3a^+a^-\cr
\lch {-2}4 &=-\fr 14\int\fr {dp}{2\pi}p_0(p_0-p)^2\left(a^+\der a^-+
a^-\der a^+\right)\cr
\lch {-1}4 &=\fr 12\int \fr {dp}{2\pi}\left[p_0^2(p_0-p)
\left(a^+\der^2a^--a^-\der^2a^+\right)+m^2
\fr {5p_0-p}{4p_0^2}a^+a^-\right]\cr}
\eqn\prep
$$
etc. We are prepared now to compute the desired differential form of
$\lch {-n}{2s}$. Using the canonical anticommutation relations
$
\left\{ a^+(p),a^-(q)\right\}=2\pi\delta(p-q)
$
and eqs. \prep\ , we get:
$$
\eqalign{
[L_{-1},\psi] &=-i\der\psi, \qquad
[L_0,\psi]=-i(\bz\bder-z\der-\fr 12)\psi ,\qquad
[\lch {-3}{4},\psi] =-i\der^3\psi\cr
[\lch {-2}4,\psi] &=-\fr i2\left[(\bz\bder-z\der)+(\bz\bder-z\der-3)\right]
\der^2\psi\cr
[\lch {-1}4,\psi] &=-\fr i2\left[(\bz\bder-z\der)(\bz\bder-z\der+1)
+(\bz\bder-z\der-3)(\bz\bder-z\der-2)\right]\der\psi\cr
[\lch {-2}6,\psi] &=-\fr i2\left[(\bz\bder-z\der)_3
+(\lo-5)_3\right]\der^2\psi .\cr }
\eqn\wi
$$
Similar calculation for $\blch {-k}4$ leads to:
$$
\eqalign{
[\blch {-1}4,\psi] &=-\fr i2\left[(\lo +1)(\lo +2)+(\lo-2)(\lo-1)
\right]\bder\psi\cr}
$$
This form of the $\lch {-k}{2s}$'s allows us to make {\it a conjecture}
about the general form of all the $\lch {-k}{2s}$ ( $0\le k\le 2s-1$ ):
$$
[\lch {-k}{2s},\psi]=
{-i\over 2}\left[(\bar z\bder -z\der)_{2s-1-k}+
(\bar z\bder -z\der -2s+1)_{2s-1-k}\right]\der^k\psi ,
\eqn\gen
$$
where $(A)_p=A(A+1)\ldots (A+p-1)$.
In order to prove our conjecture we have to be able to
derive from \gen\ the integral form of $\lch {-k}{2s}$ similar to that of
the eqs. \ftheta ,\fmomtfour ,\newch and to show that the integrands are
conserved quantities. Fortunately it exists an indirect way to prove that
\gen\ are conserved charges. It is related to the answer of the following
{\it important question} we left unanswered up to now: {\it are
the conserved
charges} \ftheta\ ,\fmomtfour\ ,\newch\ {\it etc. we have constructed,
generators of symmetries of the action} \pert\ ?Let us first check whether
the simplest nontrivial charge $\lch {-2}4$ leaves invariant the action:
$$
S=\int\left( -{1\over 2}\psi\bder\psi+{1\over 2}\bar\psi\der\bar\psi+
m\bar\psi\psi\right)d^2z\equiv\int{\cal L}d^2z .
\eqn\act
$$
By using \wi\ and
$$
[\lch {-2}4 ,\bar\psi]=(\bar z\bder -z\der -{1\over2})\der^2\bar\psi
$$
one can verify easily that
$$
[\lch {-2}4 ,{\cal L}]=\der A+\bder B .
$$
Therefore $\lch {-2}{4}$ is a generator of a specific new symmetry of
\act\ . The same is true for $\blch {-2}{4}$. Together with the Lorentz
rotation $L_0$ they close an $SL(2,R)$ - algebra. One can repeat this
calculation with $\lch {-1}{4}$, $\blch {-1}{4}$, $\lch 04$,
$\lch {-2}{6}$ etc. and the result is always that these
charges commute with the action \act\ . As it becomes clear from
this discussion the proof that $\lch {-k}{2s}$ given by eq. \gen\ are
conserved charges is equivalent to the following statement:
$
[\lch {-k}{2s},S]=0 .
$
To prove it we have to make one more conjecture, namely:
$$
\eqalign{
[\lch {-k}{2s},\bar\psi]
&={-i\over 2}\left[(\bar z\bder -z\der +1)_{2s-1-k}
+(\bar z\bder -z\der -2s+2)_{2s-1-k}\right]\der^k\bar\psi\cr
0 &\le k\le 2s-1\cr}
\eqn\conjec
$$
The remaining part of the proof is a straightforward but tedious higher
derivative calculus.

To make complete our study of the conserved charges of the off - critical
Ising model we have to find the general form of the ``conjugated charges''
$\blch {-k}{2s}$. By arguments similar to the ones presented above we
arrive to the following result:
$$
\blch {-k}{2s}={1\over 2}\left[(\bar z\bder -z\der
+\bar\alpha-2s+k+2)_{2s-1-k}+(\bar z\bder-z\der +\bar\alpha+k+1)_{2s-1-k}
\right]\bder^k,
\eqn\blss
$$
where
$\bar\alpha =-1\quad for \quad \psi$ and
$\bar\alpha =0 \quad for \quad \bar\psi$.

Our claim is that \conjec and \blss\ {\it do exhaust all the local
symmetries} (i.e. local conserved charges) of the action \act\ .
An {\it important observation} concerning the origin of these
symmetries is in order: denoting the Poincare algebra generators by
$$
L_{-1}=\der ,\qquad \bar L_{-1}=\bder ,\qquad L_0=\bar z\bder-
z\der-{1\over 2},
$$
it is obvious that the conserved charges \conjec\ and \blss\
$$
\eqalign{
\lch {-k}{2s} &={1\over 2}\left[(L_0+{1\over 2})_{2s-1-k}+
(L_0-2s+\fr 32 )_{2s-1-k}\right]L_{-1}^k\cr
\blch {-k}{2s} &= \fr 12 \left[(L_0-2s+k+\fr 32)_{2s-1-k}+
(L_0+\fr 12 +k)_{2s-1-k}\right]\bar L_{-1}^k\cr }
$$
($0 \le k\le 2s-1$),
span a specific subalgebra in the enveloping of the Poincare algebra:
$$
{\cal EP}=\left\{ L_0^{l_1}L_{-1}^{m_1}, L_0^{l_2}\bar L_{-1}^{m_2} |
L_{-1}\bar L_{-1}\sim m^2I\right\}.
$$
The condition that single out this subalgebra is the invariance of the
action \act\ . The open question, however, is about the algebraic
meaning
of such a condition. The appearence of the ${\cal EP}$ - algebra is not a
specific property of the massive Majorana fermion. Studing the conformal
limit of $\lch {-k}{2s}$ we have realized that the ``conformal''
$W_\infty$ algebra has a specific subalgebra ${\cal P}W_\infty(V)$
spanned by:
$$
\eqalign{
{\cal L}_{-k}^{(2s)} &=\fr 12\left[(\tilde L_0-\fr 12)_{2s-1-k}+
(\tilde L_0+2s-\fr 32)_{2s-1-k}\right]L_{-1}^k\cr
0 &\le k\le 2s-1,\qquad \tilde L_0=z\der+\fr 12 \cr} ,
$$
which is a subalgebra of ${\cal EP}$. All the other generators
${\cal L}_{-k}^{(2s)}$ ($k>2s-1$ and $k<0$) of $W_\infty$, however,
do not
belong to ${\cal EP}$. Considering $W_\infty$ as the largest symmetry of
the corresponding conformal model one can speculate that the only
symmetries belonging to ${\cal EP}$ survive after the perturbation.

\chapter {$W_\infty (V)$ - algebra}

Remember that our original motivation was to construct {\it noncommuting}
conserved charges for certain IM's. Having at hand the explicit form of
\conjec\ ,\blss\ of $\lch {-k}{2s}$ and $\blch {-k}{2s}$ we are prepared to
compute their algebra. As we have already mentioned $\lch {-2}4$,
$\lch {-1}4$ and $L_0$ close an $SL(2,R)$ algebra. Two more $SL(2,R)$
algebras are spanned by $\bar L_{-1}$, $\lch {-1}4$, $L_0$ and $L_{-1}$,
$\blch {-1}4$, $L_0$.
Passing to the general case let us first try to find the
structure of the ``left'' algebra, i.e.:
$$
\left[\lch {-k_1}{2s_1},\lch {-k_2}{2s_2}\right]=
\sum_{r=1}^{s_1+s_2-1}g^{s_1s_2}_{2r}(k_1,k_2)L_{-k_1-k_2}^{2(s_1+s_2-r)}.
\eqn\lef
$$
The simplest way to prove \lef\ and to compute the structure constants
$g^{s_1s_2}_{2r}(k_1,k_2)$ is based on the following ``conformal''
decomposition of the generators $\lch {-k}{2s}$ in terms of the conformal
generators ${\cal L}_{-k}^{(2s)}$:
$$
\eqalign{
\lkts  &=\sum_{l=0}^{2s-1-k}\left(\matrix{ &2s-1-k\cr &l\cr}\right)
(\bz\bder^2+\alpha\bder)^l{\cal L}^{(2s)}_{-k-l}(-m^2)^{-l} \cr
\alpha &=0\quad for\quad\psi\quad and \quad\alpha=1\quad for\quad
\bar\psi .\cr}
\eqn\decomp
$$
The fact that the operators
$
S_l=(\bz\bder^2+\alpha\bder)^l=(\bz\bder +\alpha)_l\bder^l
$
are commuting, i.e. $[S_{l_1},S_{l_2}]=0$ reduces the computation of the
structure constants $g^{s_1s_2}_{2r}(k_1,k_2)$ to the conformal ones
$C^{s_1s_2}_{2r}(k_1,k_2)$ ($k_i\le 2s_i-1$):
$$
\left[{\cal L}^{(2s_1)}_{-k_1},{\cal L}^{(2s_2)}_{-k_2}\right]=
\sum_{r=1}^{s_1+s_2-1}C_{2r}^{s_1s_2}(-k_1,-k_2){\cal
L}_{-k_1-k_2}^{2(s_1+s_2-r)} .
\eqn\cstr
$$
Note that \cstr\ is a {\cal P}$W_\infty$
subalgebra of the $W_\infty$ - algebra
\REF\pope{C.N. Pope, X. Shen, L.J. Romans\journal
Nucl.Phys.&B339(90)191.}
[\pope]
written in a specific basis of nonquasiprimary $T_{2s}$ we are using.
The remaining part of the proof that
$$
\eqalign{
g^{s_1s_2}_{2r}(k_1,k_2) &=C^{s_1s_2}_{2r}(-k_1,-k_2)\cr
0\le &k_i\le 2s_i-1\cr}
$$
is based on the following property of the conformal structure constants
\REF\ss{G. Sotkov, M. Stanishkov,{\it Infinite symmetries of the 2-D
integrable models}, preprint IFT - P.002/93}
[\ss]:
$$
\eqalign{
C^{s_1s_2}_{2r} &(-k_1,-k_2) \left(\matrix{ &2(s_1+s_2-r)-k_1-k_2-n-1\cr
&n\cr}\right)=\cr
&=\sum_l\left(\matrix{&2s_1-k_1-1\cr
&l\cr}\right)\left(\matrix{ &2s_2-k_2-1\cr &n-l\cr}\right)
C^{s_1s_2}_{2r}(-k_1-l,-k_2-n+l) .\cr }
$$
The identical statement holds for the algebra of $\blch {-k}{2s}$'s
as well. The
conclusion is that the algebra we are looking for has as subalgebras two
incomplete

\noindent
($0\le k\le 2s-1$) $W_\infty$ algebras wich do not commute
between themselves.

The general strucrure of the remaining ``left - right'' commutators:
$$
\left[ \lkts ,\blch {-l}{2p} \right]=\sum_{r=0}^{s+p-k-2}\bar g_{2r}^{sp}
(k,l)(m^2)^k\bar L_{k-l}^{2(s+p-k-1-r)}
\eqn\lrcom
$$
(if $k<l$!) is a consequence of the explicit form \gen\ and \blss\ of the
generators. In order to calculate $\bar g_{2r}^{sp}(k,l)$ we first commute
$L_{-1}^k$ and $\bar L_{-1}^l$ to the right
and then expand the both sides of
\lrcom\ in powers of $X=L_0+\fr 12$. In doing this we have to know the
coefficients $B_m^{N,a}$ in the power expancion of $(X+a)_{N+1}$:
$$
\left(X+a\right)_{N+1}=\sum_{m=0}^{N+1}B_m^{N,a}X^m .
$$
A simple combinatorial analysis [\ss] leads to the following form of
$B_m^{N,a}$:
$$
B_m^{N,a}=\fr 1{3.2^{m+1}}(N+2-m)_mA_m^{N,a}(N+2a) .
\eqn\bmna
$$
The $A_m^{N,a}$ are certain polynomials of $w=N+2a$ of degree m satisfying
the following recursion relations [\ss]:
$$
(N+2-m)_mA_m^{N,a}(N+2a)=2\sum_{p=0}^{N-m+1}(a+p)(N-p+2-m)_{m-1}
A_{m-1}^{N-p-1,a+p+1}(N+2a+p+1)
$$
and the differential equation:
$$
\fr d{dw} A_m^{N,a}(w)=A_{m-1}^{N,a}(w) .
$$
The general solution is a specific linear combination of the Bernuli
polynomials. The first few are of the form:
$$
\eqalign{
A_1^{N,a}(w) &=6w ,\qquad\qquad A_2^{N,a}(w)=3w^2-(N+2)\cr
A_3^{N,a}(w) &=w^3-(N+2)w ,\qquad A_4^{N,a}(w)=\fr {w^4}4 -\fr {N+2}2 w^2
+\fr {N^2}{N+1} \cr }.
$$
The l.h.s. of \lrcom\ contains 8 terms of the form:
$$
(X+a)_{N+1}(X+a)_{M+1}=\sum_{k+0}^{N+M+2}Y_R^{(N+M+2-k)}(aN|bM)X^k ,
$$
where
$$
Y_L^{(m)}(aN|bM)=\sum B_k^{a,N}B_{m-k}^{b,M} .
$$
Denote by $Y_L^{(m)}$ the sum of the contributions of all the 8 terms in
the l.h.s. The same expancion in the r.h.s. gives:
$$
\eqalign{
(X+c)_{M+N-2r+1} &+(X+d)_{M+N-2r+1}=\sum_m  Y_R^{(m),r}X^m\cr
Y_R^{(m),r} &=B_m^{c,N+M-2r}+B_m^{d,N+M-2r}\cr } .
$$
Then one can easily derive the following recursive relations for the
structure constants.:
$$
Y_L^{(2n-1)}=2\sum_{r=0}^{n-1}\bar g_{2r}^{sp}(k|l)Y_R^{(2n-2-2r)}
\eqn\strc
$$
The first coefficient is quite simple:
$$
\bar g_0^{sp}(k|l)=-k(2p-l-1)-l(2s-k-1) ,
$$
but the next two $\bar g_2$ and $\bar g_4$ are complicated enough
and their
explicit form does not help us in guessing the form of $\bar g_{2r}$.
We shall give here one example of the mixed left-right commutators:
$$
\eqalign{
\bigg[ \lch {-2p+2}{2p} &,\blch {-2s+3}{2s}\bigg]=(m^2)^{2(p-1)}\bigg\{
-2(2p+s-\fr 72)\bar L_{-2(s-p+1)+3}^{2(s-p+1)}+\cr
&+4(p-1)[(s-\fr 12)(-s+\fr 32)+(s-p+1)(s-p)]\bar L_{-2(s-p)+1}^{2(s-p)}
\bigg\}\cr }
$$
for $s>p$.

The remarkable observation by Thacker and Itoyama
[\ti] that the off-critical
$XY$- model has as dynamical symmetries two different Virasoro algebras is
a hint to look for Virasoro algebras generated by specific combinations of
$\lkts$ and $\blkts$. We have even an indication where to look for. It is
the fact that $\{ \fr 1{m^2}\lch {-2}4, L_0,
\fr 1{m^2}\blch {-2}4\}$, $\{\lch {-1}4,
L_0, \fr 1{m^2}\bar L_{-1}\}$ and $\{\blch {-1}4, L_0, \fr 1{m^2}L_{-1}\}$
are generators of three different $SL(2,R)$ algebras. To begin with the
second one. We have to find the analog of the ${\cal L}_2$ - Virasoro
generator (see eq. \vir\ ). An appropriate candidate for this role is
$\cL _2=\lch {-2}6+\alpha\lch {-2}4$. Simple computations lead us to the
conclusion that $\cL_{-1}=1/m^2\bar L_{-1}$, $\cL_0=L_0\equiv \lo -1/2$ and
$$
\eqalign{
\cL_1\psi &=L_{-1}^{(4)}
-\fr 94 L_{-1}=(\lo -\fr 12)(\lo -\fr 32)\der\psi\cr
\cL_2\psi &=\lch {-2}6 -3(\fr 52)^2\lch {-2}4=(\lo-\fr 12)(\lo -\fr 32)
(\lo -\fr 52)\der^2\psi .\cr}
\eqn\loat
$$
generate the incomplete Virasoro algebra $V=\{\cL_n, n\ge -1\}$. The form
\loat\ of $\cL_1$ and $\cL_2$ is very suggestive. One can easily verify
that$\cL_n$ given by:
$$
\eqalign{
\cL_n &=[\lo -\fr 12]_{n+1}\der^n ,\qquad n\ge -1\cr
[A]_k &=A(A-1)\ldots(A-k+1) ,\qquad [A]_0=1\cr }
\eqn\elen
$$
(note that $\der^{-1}\psi =-1/m^2\bder \psi$) satisfy \vir\ . One more
incomplete Virasoro algebra $\bar V$ is generated by:
$$
\bar\cL_n=[\lo +\fr 12]_{n+1}\bder^n ,\qquad n\ge -1 .
\eqn\belen
$$
The third Virasoro algebra $V_c$ spanned by:
$$
\eqalign{
\fr {(-m^2)^{1-s}}2 \lch {-2s+2}{2s} &=\fr {(-m^2)^{1-s}}2
(\lo -\fr {2s-1}2)\der^{2s-2}
\equiv\fr
{(-m^2)^{1-s}}2(L_0-s+1)L_{-1}^{2s-2}\cr
\fr {(-m^2)^{1-s}}2 \blch {-2s+2}{2s} &=\fr {(-m^2)^{1-s}}2
(\lo +\fr {2s-3}2 )\bder^{2s-2}
\equiv
\fr {(-m^2)^{1-s}}2 (L_0+s-1)\bar L_{-1}^{2s-2}\cr }
\eqn\thvir
$$
$ s=1,2,\ldots$,
plays in our opinion the major role for the exact integrability of the
$S_{13}(p)$- class of IM's. Using once more the formal identity $\bar L_{-
1}=-m^2(L_{-1})^{-1}$ we can rewrite the $V_c$ - generators \thvir\ in an
unique formula:
$$
V_n=\fr 12(-m^2)^n(L_0-n)L_{-1}^{2n} ,\qquad -\infty\le n\le\infty .
$$
As in the case of the massive Dirac fermion [\ti] one is expecting that
$V_c$ has nonzero central charge. One could see it calculating the
commutator
$$
\left[ \lch {-4}6 ,\blch {-4}6\right]
$$
and taking care about the right normal ordering of the $a^{\pm}$ operators:
$$
\eqalign{
\lch {-4}6 &=\int\fr {dq}{2\pi}q_0(q_0-q)^4(a^+\der a^-+a^-\der a^+)\cr
\blch {-4}6 &=\int\fr {dq}{2\pi} q_0(q_0+q)^4(a^+\der a^-+a^-\der a^+)\cr }
$$
The result is:
$$
\left[ \lch {-4}6 ,\blch {-4}6\right]=-8m^8L_0+m^8 .
\eqn\cch
$$
Comparing \thvir\ and \cch\ with \vir\ one concludes that:
$$
c=\fr 12 ,
$$
i.e. the massive Majorana fermion has the same central charge as the
massless one. This is in agreement with the Thacker and Itoyama's result $c=
1$ for the Dirac fermions.

This fact allows us to apply the all well- established technology of the
highest weight representations, null- vectors etc. to the case of the IM's.

\chapter {Off - critical Ward identities}

An important consequence of the fact that $\lch {-k}{2s}$ and $\blkts$ are
generators of the symmetries of the action \act\ is the following infinite
set of Ward identities for the $n$ - point Green functions of $\psi(z,\bz)$
and $\bar\psi(z,\bz)$:
$$
\eqalign{
 &\left\langle 0\left | \lkts
\Pi_{i=1}^M\psi(z_i,\bz_i)\Pi_{j=1}^N\bar\psi(z_j,\bz_j)\right |
0\right\rangle =0\cr
 &\left\langle 0\left |
\Pi_{i=1}^M\psi(z_i,\bz_i)\Pi_{j=1}^N\bar\psi(z_j,\bz_j)\blkts\right |
0\right\rangle =0\cr}.
\eqn\grf
$$
The condition for the invariance of the vacuum:
$
\lkts |0\rangle =0=\langle 0|\blkts
$
together with eqs. \gen\ ,
\conjec\ and \blss\ lead to the following system of
differential equations for $G_{MN}(z_l,\bz_l)$:
$$
\eqalign{
\bigg\{ &\sum _{i=1}^M\left[ (\loi )_{2s-1-k} +(\loi
-2s+1)_{2s-1-k}\right] \der _i^k+\cr
+  &\sum _{j=1}^N\left[ (\loi +1)_{2s-1-k} +(\loi
-2s+2)_{2s-1-k}\right] \der _j^k\bigg\}  G_{MN}(z_l,\bz _l)=0\cr }
\eqn\gmn
$$
A similar set of equations can be obtained
from the condition of $\blkts$ - symmetry of
$G_{MN}$.
Restricting ourselves to the case of 2 - point functions ($M+N=2$) we are
going to demonstrate that the Poincare invariance ($L_{-1}$, $L_0$,
$\bar L_{-1}$) and the new $SL(2,R)$ symmetries ($\lch {-2}4$,
$\blch {-2}4$, $L_0$) are sufficient to fix uniquely $G_{20}$, $G_{02}$ and
$G_{11}$ - functions. The relativistic invariance requires:
$$
G_{20} =m\sqrt {\fr {\bz}z }g_{20}(y) ,\quad
G_{02} =m\sqrt {\fr z{\bz} }g_{02}(y) ,\quad
G_{11} =img_{11}(y),\quad y=m\sqrt {-4z\bz} .
$$
The condition of $\lch {-2}4$ - invariance of $G_{20}$ leads to the
following third order differential equation:
$$
y^3g_{20}^{'''}+2y^2g_{20}^{''}-y(y^2+1)g_{20}^{'}-(y^2+1)g_{20}=0 .
\eqn\dif
$$
It happens that one can solve \dif\ in terms of $K_1(y)$ - Bessel function.
This reflects the fact that \dif\ can be obtained as a consequence of the
$K_1$ - Bessel equation:
$$
y^2g''_{20}+yg'_{20}-(y^2+1)g_{20}=0
$$
and a specific third order equation:
$$
y^3g'''_{20}-y(y^2+3)g'_{20}+(y^2+3)g_{20}=0 .
\eqn\bess
$$
The eq. \bess\ follows from the standard recursive relations for
$K_{\pm 1}$, $K_0$ and $K_2$ - Bessel functions. The $\lch {-2}4$ - Ward
identity imposes the eq. \bess\ only. Repeating the same analysis for
$G_{02}$ and $G_{11}$ we find that
$
g_{02}(y)=K_1(y)
$
and that $g_{11}$ satisfy the $K_0$ - Bessel equation:
$$
yg''_{11}+g'_{11}-yg_{11}=0 ,
$$
i.e. $g_{11}=K_0(y)$.

To make
complete our discussion of the off-critical Ising model we have to
mention that as in the conformal case the WI's \grf\ ,\gmn\ are fixing
uniquely the 2- and 3-point functions only. The calculation of, say, the 4-
point function (~using only the symmetries of the model ) requires more
information about the representations of the algebra \lef\ ,\lrcom\
we are using.
One could expect that the null-vector conditions
for the off-critical
Virasoro algebra spanned by $\lch {-2s+2}{2s}$, $L_0$ and
$\blch {-2s+2}{2s}$ will be sufficient to fix uniquely the corresponding 4-
point functions ($M+N=4$).

Continuing this line of arguments it is interesting to derive the
$\tau$ - function equation for the 2- point function $\langle\sigma
\sigma\rangle$ of the ``order parameter'' field $\sigma(z,\bz)$
\REF\mcoy{E. Barouch, B.M. McCoy, T.T. Wu\journal
Phys.Rev.Lett.&31(73)1409; T.T. Wu, B.M. McCoy, C.A. Tracy, E. Barouch
\journal Phys.Rev.&B13(76)316.}
\REF\miwa{M. Sato, T. Miwa, M. Jimbo,{\it Aspects of holonomic
quantum fields isomonodromic deformation and Ising model},
in Complex Analysis, Microlocal Calculus and Relativistic Quantum
Theory, D. Iagolnitzer (ed.), Lecture Notes in Physics, vol.126,
Berlin, Heidelberg, New York, Springer.}
[\mcoy,\miwa] from the corresponding WI's for $\sigma$ and the null-
vector conditions  .
One could find the infinitesimal $\lch {-2}4$ transformation of
$\sigma$ using the explicit realization of $\sigma$ in terms of $\psi$
and $\bar\psi$
[\miwa] or the conformal perturbation technics. The
result is quite surprizing. Following the analogy with the conformal
WI's one is expecting the commutator $[\lch {-2}4 ,\sigma(z,\bz)]$ to
have a form similar to \wi\ for $\psi$ with some differencies in the
coefficients of the differential operators. However, the difference
with the $[\lch {-2}4,\psi]$ is more drastic: the new field
$\sigma_3(z,\bz)$ wich is a specific descendent of $\sigma$ coming
from the off-critical OPE $T(z,\bz)\sigma(w,\bar w)$ contributes to
the commutator [\ss]. This fact makes the corresponding WI's much more
complicated due to the presence of the new 2-point function
$\langle\sigma_3\sigma\rangle$ together with
$\langle\sigma\sigma\rangle$.

It is important to note that the simple {\it differential form}
 \gen\ ,
\conjec\ ,\blss\ {\it of the} $\lkts$ {\it is a specific property of
the free massive fields only} (fermions or bosons). As we shall
demonstrate in the case of the off-critical Potts model the action of
the $\lkts$ - symmetry (even in its simple $(SL(2,R)$ part) on {\it
the interacting fields always involve specific descendents of these
fields}.

\chapter {Potts model}

Our discussion up to now was concentrated on the off-critical
symmetries of the Ising model. The goal was to demonstrate on the
simplest example how one can construct new conserved charges , how to
compute their algebra and how to use this algebra in the calculations
of the correlation functions. It is almost evident that one can
generalize all these constructions for arbitrary number of free
massive fermions and bosons belonging to different representations of
$O(n)$ (or $SU(n)$)
\REF\aass{E. Abdalla, M.C.B. Abdalla, G. Sotkov, M. Stanishkov,{\it
Off - critical current algebras}, in preparation.}
[\aass]. The true question however is how one can
realize all this program of {\it describing the IM's in terms of the
representations of their off-critical symmetry algebra} in the case of
interacting fields. Turning back to the set of the IM's known as
$\Phi_{13}$ - perturbations of the minimal models, we shall consider
the models with
{\it even} $p$ ( $c_p=1-\fr 6{(p+1)(p+2)}$ ) only. The
reason is that in this case the $T_{2s}$ conservation laws has a
specific term [\zam] we need in our constructions. The first such
model ($p=2$) is the thermal perturbation of the Ising model. The next
is the so called ``kinks perturbation'' of the 3- state Potts model
\REF\fat{V. Fateev\journal Int.Journ.Mod.Phys.&A2(91)2109.}
[\fat]:
$$
S_{13}(p=4)=S_{Potts}(c=4/5)+g\int\left(\psi_{2/3}\bar\psi_{2/3}^++
\psi_{2/3}^+\bar\psi_{2/3}\right)d^2z .
$$
In order to derive the specific form of the conservation laws for,
say, $T_4$ and $T_6$, we cannot use anymore the free massive fermions
technics described above. What we can do here ( and for each of the
models of this class) is: 1)to perform the perturbation expansion
around the conformal point ($g=0$); 2) to use at each order of the
perturbation series the conformal WI's and at the end 3) to resum the
perturbation series. To begin with the $T_4$ conservation law. We have
to calculate the mean value of $\bder T_4$ in the presence of an
arbitrary set of other fields:
$$
\eqalign{
\bder\left\langle T_4(z,\bz)\ldots\right\rangle &=\bder\left\langle
T_4(z)\ldots\right\rangle_{conf}+g\bder\int d^2w\left\langle
T_4(z)\psi_{2/3}(w)\bar\psi_{2/3}^+(\bar w)\ldots\right\rangle+\cr
&+g^2\bder\int d^2w_1\int d^2w_2\left\langle
T_4(z)\psi_{2/3}(w_1)\bar\psi_{2/3}^+(\bar
w_1)\psi_{2/3}(w_2)\bar\psi_{2/3}^+(\bar
w_2)\ldots\right\rangle+\ldots\cr}.
$$
Using the conformal OPE's:
$$
T_4(z)\psi_{2/3}(0)=\left\{\fr {\alpha_0}{z^4}+\fr
{\alpha_1}{z^3}\der+\fr 1{z^2}(\alpha_2L_{-2}+\alpha_3\der^2)+\fr
1z(L_{-3}+\alpha_4\der
L_{-2}+\alpha_5\der^3)\right\}\psi_{2/3}(0)+\ldots
$$
the level-3 null vector
$$
\left( L_{-3}+\gamma_1L_{-1}L_{-2}+\gamma_2L_{-1}^3\right)\psi_{2/3}=0
$$
one can get read of $T_4$ at each order in $g$. The next step is to
take some of the integrals and then to resum the perturbation series.
The exact result is:
$$
\left\langle\bder T_4\ldots\right\rangle=\left\langle\left(
a\der^3\Theta+b\der L_{-2}\Theta+cg^3\der T\right)\ldots\right\rangle
\eqn\pottf
$$
where $\Theta$ is the trace of the stress-tensor:
$$
\bder T=\der \Theta ,\qquad \Theta=\fr g3
(\bar\psi^+_{2/3}\psi_{2/3}+\bar\psi_{2/3}\psi^+_{2/3})
$$
and a, b, c are fixed numerical constants. Since $\psi_{2/3}$ as a
conformal field has no null vector at the second level:
$$
L_{-2}\psi_{2/3}\not= \fr 9{14}\der^2\psi_{2/3}
$$
the form \pottf\ of the $T_4$ conservation law is not the desired one
\tfour\ . According to our criterion \claws\ one can naively conclude
that this model has no any nontrivial noncommuting conserved charges.
However such a conclusion is wrong and the reason is in the broken
$W_3$ - symmetry of the model. Using the conformal OPE of the $W_3$
current
\REF\zf{
V. Fateev, A.B. Zamolodchikov\journal Nucl.Phys.&B280[FS18](87)644.}
[\zf]:
$$
\eqalign{
W_3(z)\psi_{2/3}(0) &=\left\{\fr {w_0}{z^3}+\fr 1{z^2}W_{-1}+\fr 1z
W_{-2}+\ldots\right\}\psi_{2/3}\cr
W_n &=\int z^{n+2}W_3(z)dz ,\qquad w_0=\fr 29\sqrt {\fr {26}{15}}\cr}
$$
one can apply the method of the perturbative expansion described above
and to prove that $W_3$ is not conserved:
$$
\bder W_3=w_0\der^2\Theta+\alpha\der W_{-1}\Theta +\beta W_{-2}\Theta
\eqn\wtre
$$
due to the last term in the r.h.s. The crucial point is that the field
$\psi_{2/3}$ satisfies the following specific null vector conditions
of the bigger $W_3$ algebra:
$$
\left( W_{-1} -\fr 12 \sqrt{\fr {26}{15}}L_{-1}\right)\psi_{2/3}=0 ,\quad
\left( W_{-2} -\fr 6{13} \sqrt{\fr {26}{15}}\left( 2L_{-1}^2-\fr 53
L_{-2}\right)\right)\psi_{2/3}=0 .
$$
Therefore we can rewrite \wtre\ in the following simple form:
$$
\bder W_3=A\der^2\Theta +BL_{-2}\Theta .
\eqn\dw
$$
Combining eqs. \dw\ and \pottf\ we arrive at the desired spin-4
conservation law:
$$
\bder \left(T_4-\fr bB\der W_3\right)=a\der^3\Theta +cg^3\der T .
\eqn\spfor
$$
Now we are able
to construct five new conservation laws. The simplest one
is $\bder {\cal F}_4^{(1)} =\der\Theta_4^{(1)}$
$$
{\cal F}_4^{(1)} =zT_4-z\fr bB\der W_3+bg^3\bz T ,\quad
\Theta_4^{(1)} =a\der^2(z\Theta)-3a\der\Theta+g^3bzT+abg^4\bz
\Theta
$$
and the corresponding conserved charge has the form:
$$
\lch {-2}4=\int{\cal F}_4^{(1)}dz-\int\Theta_4^{(1)}d\bz .
$$
Comparing eqs. \spfor\ and \tfour\ we can say that all the charges
$\lch {-k}4$, $\blch {-k}4$ ($k=0,1,2$) can be obtained from the ones
of the Ising model by substituting $T_4\rightarrow T_4-\fr bB \der
W_3$, $m^2\rightarrow g^3$, $\beta\rightarrow c$ etc.

The proof that the $T_6$ conservation law has the form \newcl\ is more
involved [\ss]
and it requires together with the $W_3$ - nonconservation law
\dw\ an analogous nonconservation law for the spin-5 current $W_5\sim
:TW_3:-\alpha\der^2W_3$:
$$
\bder W_5=\beta_1\der^4\Theta+\beta_2\der^2 L_{-2}\Theta+\beta_3
L_{-4}\Theta+\beta_4 L_{-2}^2\Theta .
$$
Crucial for these constructions is the fact that in this case we
have three independent spin-6 tensors, i.e. $T_6^{(1)}\sim T^3$,
$T_6^{(2)}\sim (\der T)^2$ and $T_6^{(3)}\sim W_3^2$.

Having constructed the conserved charges $\lkts$ ($0\le k\le 2s-1$) we can
compute their algebra using the conformal perturbation technics. The
simplest but highly nontrivial calculation is the one of the commutators of
$\lch {-2}4$, $\blch {-2}4$ and $L_0$. The surprising fact is that they
close again a $SL(2,R)$ algebra as in the case of the Ising model.
This is
an indication that {\it the conserved charges of the off-critical Potts
model could have the same algebra \lef\ ,\lrcom\
as the charges of the Ising model}.
Accepting this as an working hipothesis we have to mention the crucial
difference between the off-critical fermion and parafermion. Although they
have the common algebra of symmetries the generators of these symmetries
$\lkts$ are acting on $\psi$ and $\psi_{2/3}$ in a completely different
way:
$$
\eqalign{
\left[ \lch {-3}{4},\psi_{2/3}(z,\bz)\right] &=:\left( \fr 32\der^3+
\fr 3{20}\der T(z,\bz)\right) \psi(z,\bz):\cr
\left[ \lch {-2}4,\psi_{2/3}(z,\bz) \right] &=:\left( g^3\bz\der+(
\fr 45 \der^2+
\fr 32 T)+z(\fr 3{2}\der^3+\fr 3{20}\der T)\right) \psi(z,\bz):\cr
\left[ \lch {-1}4,\psi_{2/3}(z,\bz)\right] &=:\bigg( g^3(\fr 43 \bz\der +2z
\bz\der -\bz^2\bder)+\fr {13}{10}\der+2z(
\fr 45\der^2+\fr 32 T)+\cr  &+z^2(\fr 32\der^3+
\fr 3{20}\der T)\bigg) \psi(z,\bz):\cr }
\eqn\powi
$$
etc. The corresponding commutators of $\lch {-k}6$ with $\psi_{2/3}$
contain terms with $T_4$ and $T$ and their derivatives and so on. Therefore
the corresponding WI's equations relate in fact the functions of the
parafermions with the functions of their descendents.
As one can see from the conformal limit $g\rightarrow 0$ this complicated
form of the WI's is not a specific feature of the off-critical model only.
The WI's for the conformal $CW_\infty$ also include the descendents of the
parafermion.

\chapter {Few conjectures}

We are prepared now for {\it our main conjecture} concerning the symmetries
of the IM's given by the action:
$$
S_{13}(p=even)=S_{conf}(c_p)+g\int
\Phi_{13}(z,\bz)\bar\Phi_{13}(z,\bz)d^2z .
$$

{\it All these models have} $W_\infty (V)$ \lef\ ,\lrcom\
{\it as a symmetry algebra and they
belong to different representations of this algebra}.

One could try to prove it styding model by model the $S_{13}$ (p=even) set
of IM's in order to understand the general properties of the
representations of the $W_\infty$ - algebra. There are many signs that for
the practical purposes such as the computation of the correlation functions
it is sufficient to classify and explicitely construct the representations
of the one of the Virasoro subalgebras of $W_\infty(V)$,
namely the one spanned
by $\lch {-2s+2}{2s}$, $L_0$ and $\blch {-2s+2}{2s}$.

The set of models given by $S_{13}$ (p=even) represents only a first member
of the infinite family of relativistic IM's. What one could expect for the
other members of this family? Could we realize them as
specific
representations of certain infinite algebras and how these algebras
look like? Our hipothesis for the symmetries of the IM's is based on their
classification according to the admissible spins of the commuting conserved
charges $P_s$, $\bar P_s$. In the case of $S_{13}(p)$ series we have:
$$
s=1\qquad (mod 2) ,
$$
i.e. the Coxeter exponents modulo the Coxeter number of $SU(2)$. Their
symmetry algebra $W_\infty$ ($V$) could be considered as a specific
deformation of the ${\cal PW}_\infty$ subalgebra of $W_\infty$
generated by
the currents $T_{2s}$. Consider a set of models with $s=1 (mod 3)$ as
allowed set of spins for $P_s$, $\bar P_s$.
They are specific perturbations of the minimal models of the $W_3$ -
algebra [\fat]. Therefore the adequate conformal $W_\infty$ algebra is
generated by the descendents $T_{2p}$, $W_{2p+1}$ of the stress-tensor
$T_2$ and the spin-3 current $W_3$. Then the off-critical infinite algebra
$W_\infty$ ($W_3$) describing this series of IM's can be obtained as a
deformation of the $W_3$ - analog of ${\cal PW}_\infty$, i.e. a specific
subalgebra of the conformal $CW_\infty (W_3)$. A straightforward {\it
conjecture} is that the IM's with the spins of $P_s:s=1 (mod N)$ can be
realized as specific representations of the $W_\infty (W_N)$ - algebras.

We want to mention in conclusion that one of the important lessons we have
learned studing the off-critical properties of the Ising and Potts models
is the crucial role of the {\it conformal} ${\cal PW}_\infty$ - algebras
(and their Virasoro subalgebras) in the description of these models as
representations of certain infinite algebra $W_\infty (V)$. The
construction of the relevant {\it conformal} $W_\infty (G)$ algebras is in
our opinion the most important part of the ambicious program of
classification of the IM's according to their infinite algebra of
symmetries.
\vskip1cm

We are gratefull to FAPESP for the financial support and to our collegues
from IFT and USP for the stimulating scientific atmosfere during our stay
in Brazil.

\refout
\bye